# Multivariable-based Correlation Dimension Analysis for Generalized Space

Yanguang Chen

(Department of Geography, College of Urban and Environmental Sciences, Peking University, 100871, Beijing, China. Email: chenyg@pku.edu.cn)

**Abstract**: Fractal geometry proved to be an effective mathematical tool for exploring real geographical space based on digital maps and remote sensing images. Whether the fractal theory tool can be applied to abstract geographical space has not been reported. An abstract space can be defined by multivariable distance metrics, which is frequently met in scientific research. Based on the ideas from fractals, this paper is devoted to developing correlation dimension analysis method for generalized geographical space by means of mathematical derivation and empirical analysis. Defining a mathematical distance or statistical distance, we can construct a generalized correlation function. If the relationship between correlation function and correlation lengths follows a power law, the power exponent can be demonstrated to associate with correlation dimension. Thus fractal dimension can be employed to analyze the structure and nature of generalized geographical space. This suggests that fractal geometry can be generalized to explore scale-free abstract geographical space. The theoretical model was proved mathematically, and the analytical method was illustrated by using observational data. This research is helpful to expand the application of fractal theory in geographical analysis, and the results and conclusions can be extended to other scientific fields.

# 1 Introduction

Space concept is usually characterized by distance such as road distance and railway distance in geography. Distance decay law is one of basic law in geo-spatial analysis. The prerequisite for using distance to effectively describe spatial features is that the space has a characteristic scale. If distance decay takes on a power law, the space bear no characteristic scale, and the distance-based spatial analysis should be replaced by dimension-based spatial analysis. Fractal geometry provides a powerful tool for dimension-based spatial analysis. Geographical space include Euclidean space and fractional space. The former bears characteristic scales and can be described with conventional mathematical methods, while the latter bear no characteristic scale should be described with fractal geometry. In this sense, the fractional geo-space corresponds to the systems with fractal structure. In fact, the creation of fractal theory is related to geographical fractal phenomena such as coastlines and city-size distributions (Mandelbrot, 1965; Mandelbrot, 1982). Fractal geometry has been applied to geographical analysis for the scale-free phenomena in the real world (Batty, 2005; Batty and Longley, 1994; Chen, 2008; Frankhauser, 1994; Goodchild and Mark, 1987; Rodriguez-Iturbe and Rinaldo, 1997). However, the relevant research is mainly limited to the physical space of reality. By means of mathematical metrics, we can define generalized geographical space. Whether fractal geometry can be applied to abstract geographical space is still unknown. If we can find an approach to modeling generalized space of geographical systems, we can develop new spatial analysis methods for geography, and thus promote the application function of fractal geometry.

One of the breakthroughs in solving the problem lies in the correlation function, and the key of constructing correlation function is to define abstract distance. If we can define a proper mathematical distance or statistical distance, then we can generate an abstract geographical space and construct a spatial correlation function. If the correlation function follow scaling law, we can apply fractal geometry to characterize the abstract space of geographical systems. The abstract space can be treated a generalized geographical space. In fact, the definition of abstract distance is not a problem, because there are ready-made metric theories and methods available. The problem lies in what methods are used to reveal the scale-free distribution and self-similar structure that may exist in abstract space. This paper is devoted to developing a distance-based approach to calculating correlation dimension of generalized geographical space. The rest parts are organized as follows. In Section 2, based on metric theory, generalized geographical space is defined, spatial correlation function is constructed, and spatial correlation dimension is mathematically derived for the generalized space. In Section 3, an empirical study is made to demonstrate how to use this type of spatial correlation analysis. In Section 4, several related questions are discussed, and finally, in Section 5, the discussion is concluded by summarizing the main points of this work.

## 2 Theoretical models

### 2.1 Metric and generalize geographical space

For a geographical system, spatial correlation dimension can be defined by distance. If a distance matrix is obtained for a real geographical space, we can calculate the spatial correlation dimension (Chen, 2008; Chen and Jiang, 2010). Generalizing the distance based on the real space to the distance based on mathematical space, we can compute correlation dimension in a similar way. Thus maybe we can reveal fractal structure in a generalized spatial system. The precondition is to define a proper distance metric. Suppose there is a geographical system of $n$ elements, which can be measured by $m$ variables. Thus the observed data can be expressed as a matrix as below:

$$\mathbf{X} = \begin{bmatrix} \mathbf{x}_1 & \mathbf{x}_2 & \cdots & \mathbf{x}_m \end{bmatrix} = \begin{bmatrix} \mathbf{x}_1^* \\ \mathbf{x}_2^* \\ \vdots \\ \mathbf{x}_n^* \end{bmatrix} = \begin{bmatrix} x_{11} & x_{12} & \cdots & x_{1m} \\ x_{21} & x_{22} & \cdots & x_{2m} \\ \vdots & \vdots & \vdots & \vdots \\ x_{n1} & x_{n2} & \cdots & x_{nm} \end{bmatrix}_{n \times m} = [x_{ik}]_{n \times m}, \quad (1)$$

where $\mathbf{X}=[x_{ij}]$ denotes observational data matrix, the row vector $\mathbf{x}_i^*=[x_{j1}, x_{j2},\ldots, x_{jk},\ldots x_{jm}]$ is used to describe sample points, and the column vector $\mathbf{x}_j=[x_{i1}, x_{i2},\ldots, x_{ij},\ldots x_{in}]^T$ is used to describe variables ($i, j=1,2,\ldots,n$; $k=1,2,\ldots,m$). The sample covariance matrix of these $m$ variables is

$$\mathbf{V} = \begin{bmatrix} v_{11} & v_{12} & \cdots & v_{1m} \\ v_{21} & v_{22} & \cdots & v_{2m} \\ \vdots & \vdots & \vdots & \vdots \\ v_{m1} & v_{m2} & \cdots & v_{mm} \end{bmatrix}_{m \times m} = [v_{kl}]_{m \times m}, \quad (2)$$

where $k, l=1,2,\ldots, m$. Then we can define Euclidean distance, Minkowski distance, and statistical distance. Euclidean distance can be treated as special case of statistical distance and Minkowski distance. The well-known Euclidean geometric distance is defined as follows

$$d_{ij} = [\sum_{k=1}^{m} |x_{ik} - x_{jk}|^2]^{1/2}, \quad (3)$$

where $d_{ij}$ refers to the distance between the $i$th element and the $j$th element. Minkowski distance represents a mathematical distance, which is defined as below:

$$d_{ij}(p) = [\sum_{k=1}^{m} |x_{ik} - x_{jk}|^p]^{1/p}, \quad (4)$$

where the parameter $p=1, 2,\ldots, \infty$. If $p=1$, we have city block distance, which is also termed absolute distance. If $p=2$, equation (4) changes to equation (3) and we have Euclidean distance. If $p \to \infty$, we have Chebychev distance, $d_{ij}(\infty) = \max(|x_{ik} - x_{jk}|)$, which is also termed square distance or box distance.

An important definition of distance is what is called statistical distance. For an $m$-by-$m$ square matrix, **B**, , if quadratic form $\mathbf{x}^T \mathbf{B} \mathbf{x} > 0$ is valid for any nonzero vector **x**, then we say **B** is a positive-definite matrix. If **B** is a positive-definite matrix, a statistical distance can be defined as follows

$$d_{ij} = \left[ (\mathbf{x}_i - \mathbf{x}_j)^T \mathbf{B} (\mathbf{x}_i - \mathbf{x}_j) \right]^{1/2} . \tag{5}$$

Different structures of **B** result in different distances. If **B**=**E**, where **E** represents an identity matrix, we have Euclidean distance. If $\mathbf{B} = \mathrm{diag}(1/\sigma_1^2, 1/\sigma_2^2, \ldots, 1/\sigma_m^2)$, where $\sigma$ denotes the standard deviation of the $k$th variable, we have precisely weighted distance. If $\mathbf{B} = \mathbf{V}^{-1}$, where **V** is the sample covariance matrix of the $m$ variables, we have Mahalanobis distance (Mahalanobis, 1936). Thus, for $n$ geographical elements, we have an $m$-by-$m$ distance matrix based on $m$ variables as follows

$$\mathbf{D} = \begin{bmatrix} d_{11} & d_{12} & \cdots & d_{1n} \\ d_{21} & d_{22} & \cdots & d_{2n} \\ \vdots & \vdots & \cdots & \vdots \\ d_{n1} & d_{n2} & \cdots & d_{nn} \end{bmatrix}_{n \times n} = \left[ d_{ij} \right]_{n \times n} . \tag{6}$$

These distances are defined in a mathematical space rather than real geographical space. Based on the generalized distances, we can calculated generalized correlation dimension. In practical application, in order to minimize the influence of dimension, it is necessary to standardize or normalize the variables before calculating the distance matrix. Based on the Z-score standardized variables, the Mahalanobis distance can be proved to equal the Euclidean distance of standardized principal components or factors.

## 2.2 Preparation

The simplest correlation function can be constructed by means of step function. Suppose there is a geographical element, say, city $i$, at location $x$, the number of city is $N(x) \equiv 1$. Then, can another element, say, city $j$, be found within a certain distance from this city? If we can find another city, the result is 1, or else, the result is 0. Thus, the spatial correlation process can be converted into a distance measurement process. If the distance between city $i$ and $j$ is less than or equal to a critical length $r$, the result is 1, or else, it is 0. The critical length represents a threshold distance, which takes on a radius in practical measurement. Then, the measurement results of finding cities can be expressed by Heaviside step function based on a dummy variable (Manrubia *et al*, 1999), that is

$$N(x,r) = H(r - d_{ij}) = \begin{cases} 1, & d_{ij} \leq r \\ 0, & d_{ij} > r \end{cases}, \quad (7)$$

where *H* refers to Heaviside function, $d_{ij}$ denotes the distance between city *i* and city *j*, and $N(x, r)$ is the number of cities that we can find within the circular area with radius *r*. Heaviside function is a discontinuous function which is also called Heaviside step function (Davies, 2002; Zhang and Zhou, 2021). The radius *r* can be treated as correlation length in spatial analysis, but it represents a yardstick in fractal dimension measurement. The center of the area is location *x*. Based on a scale *r*, all cities can be divided into two categories: one is correlated to city *i*, and the other is not correlated to city *i*. According to the principle of generating correlation function, the cumulative correlation function indicating correlation sum can be calculated by

$$C(r) = \frac{1}{N^2} \sum_{i=1}^{N} \sum_{j=1}^{N} H(r - d_{ij}) = \frac{1}{N^2} \sum_{i=1}^{N} \sum_{j=1}^{N} N(x) N(x+r). \quad (8)$$

In this way, based on Heaviside function, the correlation function is expressed as a discrete form. If the focus is on *x*, then *x*=0. Equation (8) represents the general form of spatial correlation function. It is a point-point correlation function, indicating a global correlation. If *x* is fixed to a given point, equation (8) will change to a one-point correlation function, indicating a local correlation. In order to derive the relation between the correlation function for generalized space and fractal dimension, we have to consider the spatial relationships of cities in an urban network.

## 2.3 Derivation of spatial correlation dimension

Starting from Renyi entropy, we can derive the generalized spatial correlation dimension by analogy with multifractal dimension. A basic postulate is that there is power law relation between spatial correlation function *C*(*r*) and correlation length *r*. Equivalently, the relation between correlation number *N*(*r*) and correlation length *r* follows a power law. In the process of measurement, the correlation length can be represented by yardstick length. The basis of theoretical analysis is box-counting method of fractal dimension. Based on Renyi entropy (Rényi, 1970), a generalized correlation dimension is defined as follows (Feder, 1988; Hentschel and Procaccia, 1983; Halsey *et al*, 1986; Vicsek, 1989)

$$D_q = \frac{1}{q-1} \lim_{\varepsilon \to 0} \frac{\ln \sum_{i=1}^{N(\varepsilon)} P_i(\varepsilon)^q}{\ln \varepsilon}, \qquad (9)$$

where $q$ denotes the order of moments ($q=-\infty,\ldots,-2,-1,0,1,2,\ldots,\infty$), $\varepsilon$ refers to the linear size of box, $P_i$ is the ratio of the number of fractal elements appearing in the $i$th box to the number of all fractal elements, that is, $P_i=N_i(\varepsilon)/N$. If $q=2$ as given, we have correlation dimension $D_2$ (Grassberger, 1983). The correlation length can be taken as the reciprocal of the linear size of box, that is, $r=1/\varepsilon$. Let $q=2$, and $\varepsilon=1/r$, then equation (9) changes to a formula of spatial correlation dimension.

Suppose there is a geographical region with $N$ elements inside (e.g., cities and towns). The fractal dimension of the spatial distribution of these elements can be measured by box-counting method. Consider one of elements, say, the $i$th element. Taking the element as center, we can draw a circle with radius $r$. The radius $r$ can be treated as a correlation length. Investigate how many elements fall into this circle, the number can be recorded as $N_{ij}(r)$, where $i,j=1,2,3,\cdots,N$. For $N$ cities, we can draw $N$ circles. Changing the radius is equivalent to changing the correlation length, and focusing on different elements is equivalent to changing the correlation center. For the $i$th element, the correlation function based on radius $r$ is

$$C_i(r) = \frac{1}{N} \sum_{j=1}^{N} N_{ij}(r) = \frac{1}{N} \sum_{j=1}^{N} H(r-d_{ij}) = \frac{N_i(r)}{N} = P_i(r), \qquad (10)$$

where $d_{ij}$ denotes the distance between element $i$ and element $j$, the correlation length $r$ is equivalent to a length of measurement yardstick. Equation (10) gives the probability of finding other elements around element $i$. Accordingly, the total correlation number of the $i$th element is

$$N_i(r) = \sum_{j=1}^{N} N_{ij}(r) = \sum_{j=1}^{N} H(r-d_{ij}). \qquad (11)$$

As indicated above, Heaviside function is a step function as below:

$$N_{ij}(r) = H(r-d_{ij}) = \begin{cases} 1, & d_{ij} \leq r \\ 0, & d_{ij} > r \end{cases}. \qquad (12)$$

This implies that the spatial correlation is based on dummy variable. According to the property of Heaviside function, if the distance of element $j$ to element $i$ is $d_{ij}<r$, element $j$ will be taken into account, and $N_{ij}(r)=1$, otherwise it will be ignored, and $N_{ij}(r)=0$. If there are $N_i(r)$ elements fall into the field of radius $r$, then the local correlation function based on elements $i$ is the ratio of the elements within the circle of radius $r$ to all elements, that is $P_i(r)= N_i(r)/N$. In terms of the definition of correlation, the global correlation function for the $N$ element in the geographical region can be expressed as

$$C(r) = \frac{1}{N}\sum_{i=1}^{N} C_i(r) = \frac{1}{N}\sum_{i=1}^{N} P_i(r) = \frac{1}{N^2}\sum_{i=1}^{N}\sum_{j=1}^{N} H(r - d_{ij}), \quad (13)$$

which is based on equation (10). The key is to find the mathematical association of equation (13) with equation (9).

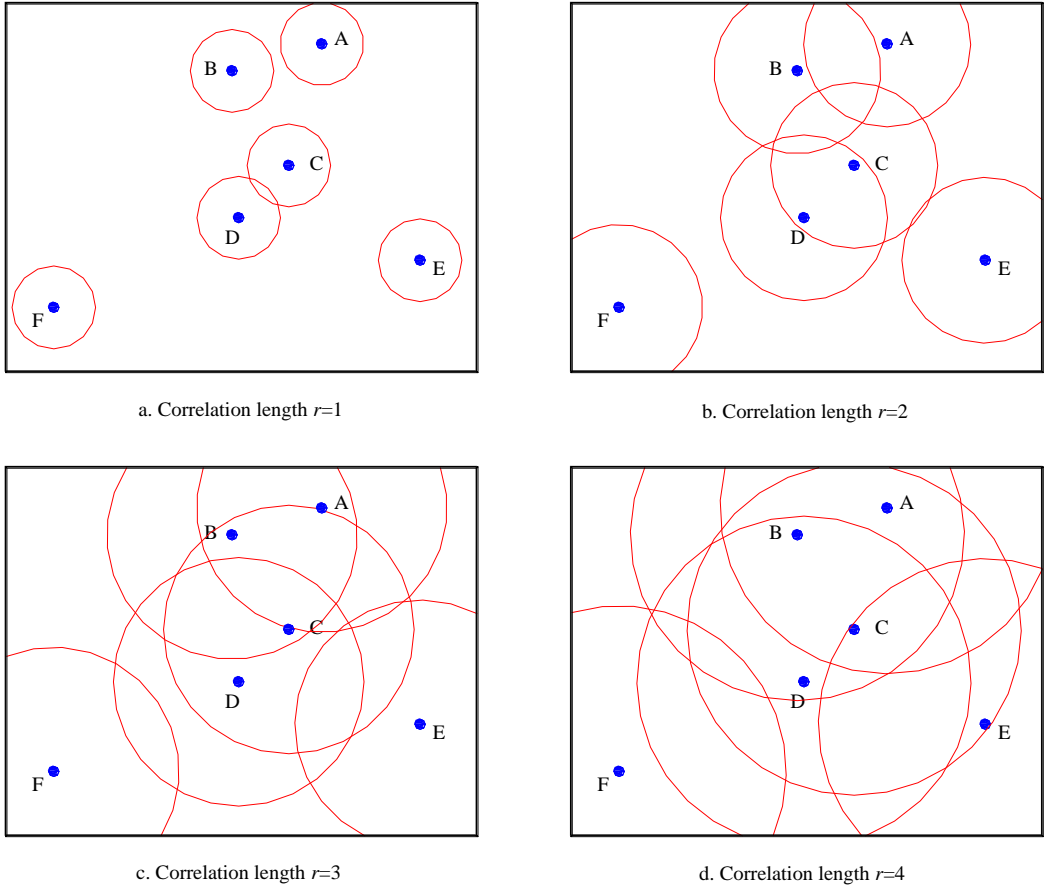

**Figure 1. A sketch map of point-point correlation in generalized geographical space**

Note: Six points represent a simple system of cities. Figure 1(a), yardstick length $r=1$, correlation number $N(r) = 6$, no inter-correlation element. Figure 1(b), yardstick length $r=2$, correlation number $N(r)=10$, four inter-correlation elements. Figure 1(c), yardstick length $r=3$, correlation number $N(r) =12$, six inter-correlation elements. Figure 1(d), yardstick length $r=4$, correlation number $N(r) =14$, eight inter-correlation elements.

It should be noted that spatial correlation is a process of interdependence of all elements in a network. A correlation process bears two characteristics. First, symmetry. Correlation is an interactive process: if city $i$ is correlated with city $j$, city $j$ is also correlated with city $i$, so the number of correlations is calculated twice. If the two-way relationships are replaced by one-way relationships, the correlation integral process will miss parts of links. Second, reflexivity. Each city is of self-correlation. If self-correlation is neglected, the box counting will miss parts of points. In this case, the spatial information will not be complete. Therefore, if $N^2$ in equation (13) is replaced by $N(N-1)$, as done in literature, the result will be inaccurate.

Now, spatial correlation process can be associated with box-counting method. Based on a correlation length $r$, all the correlated elements can form a number of sets. Different sets comprise different numbers of elements (Figure 1). Each set represents a cluster. The number of clusters is $N_b(r)$, which comes between 1 and $N$, that is, $1 \leq N_c(r) \leq N$. The number of elements in the $k$th cluster is $N_h(r)$, and the number of cluster is $h=1, 2, \ldots, N$. Clearly, the number $N_h(r)$ also comes between 1 and $N$, i.e. $1 \leq N_h(r) \leq N$. Changing the length of correlation radius $r$ yields different number of clusters, $N_c(r)$, and number of elements in each cluster, $N_h(r)$, will change. If the correlation length $r$ is very small, the number of clusters will be $N_c(r)=N$, and the number in each cluster will be $N_h(r)=1$. In this case, each element represents a cluster. On the contrary, if the correlation length $r$ is large enough, all the elements are grouped to a single cluster, and in this instance, we have $N_c(r)=1$, and $N_h(r)=N$. If we want to cover these clusters with a minimum number of boxes, we will need $N_c(r)$ boxes with linear size of $2r$. In other words, the Feret diameter of a circular box is $2r$.

Next, let's count the cumulative number of global correlation in the process of spatial measurement. According to the spatial correlation rule, if a cluster consists of $N_h(r)$ elements, the probability of local spatial correlation around element $i$ is $P_i=N_h(r)/N$, but the cumulative probability of local correlation is $P_h= N_h(r)^2/N$. Thus, equation (9) can be re-expressed as

$$C(r) = \frac{1}{N} \sum_{i=1}^{N(r)} \frac{N_i(r)}{N} = \frac{1}{N} \sum_{h=1}^{N_c(r)} N_h(r) \frac{N_h(r)}{N} = \sum_{h=1}^{N_c(r)} P_h(r)^2, \qquad (14)$$

where $P_h=N_h(r)/N$ is the element-element correlation probability based on correlation length $r$. If the spatial correlation process leads to a fractal patterns, then we have a power law as follows

$$C(r) = C_1 r^{D_c}, \qquad (15)$$

where $D_c$ denotes spatial correlation dimension, and $C_1$ is the proportionality coefficient. Theoretically, $C_1=1$. Therefore, in the process of mathematical reasoning, the coefficient $C_1$ can be ignored. Substituting equation (14) into equation (15) yields

$$D_c = \lim_{r \to \infty} \frac{\ln \sum_{k=1}^{N_c(r)} P_k(r)^2}{\ln(r)}. \qquad (16)$$

By comparison, we can find that equation (16) is in fact a special case of equation (9). Let $q=2$, and $r=1/\varepsilon$. Then, equation (9) changes to equation (16). The difference lies in that equation (16) is based on a power law, while equation (9) is based on an inverse power law. Based on point-point correlation function, equation (15) gives a point-point correlation dimension, indicating a global correlation dimension. If the point-point correlation function is reduced to a one-point correlation function, the point-point correlation dimension will change to one-point correlation dimension, indicating a local correlation dimension. The one-point correlation dimension is similar to fractal mass dimension (Sambrook and Voss, 2001), or radial dimension (Frankhauser, 1998a).

**2.4 Bounds of generalized spatial correlation dimension**

It can be proved that the value range of generalized spatial correlation dimension is between 0 and 2. If all the elements are concentrated at one point, we will have $D_c=0$. If all the elements are evenly distributed in the whole region, the dimension is $D_c=2$. If there is only one element such as a city in a geographical region, the distribution of will be extremely concentrated, and no other correlated element can be found in the region. In this case, the element number is $N(r)=1$, the probability measure is $P_k=P=1$, the linear size of box approaches to zero. Thus we have

$$D_c^{(l)} = \lim_{r \to \infty} \frac{\ln \sum_{k}^{N(r)} (1/N(r))^2}{\ln(r)} = -\frac{\ln((1/N(r)^2)}{\ln r} = \frac{\ln(1)}{\ln r} = 0, \qquad (17)$$

which is just the topological dimension of fractal point sets. On the contrary, if all the elements fit evenly over the area, the probability measure is $P_k=P=1/N(r)$, and the linear size of box is $r=1/N(r)^{1/2}$, thus we have

$$D_c^{(u)} = \frac{\ln \sum_{k}^{N(r)} (1/N(r))^2}{\ln r} = \frac{\ln[N(r)(1/N(r))^2]}{\ln(1/\sqrt{N(r)})} = \frac{2\ln(1/N(r))}{\ln(1/N(r))} = 2, \qquad (18)$$

which is just the Euclidean dimension of the embedding space of a fractal object defined the generalized geographical region. In equations (17) and (18), the letters "l" and "u" denote "lower" and "upper", respectively. The spatial correlation dimension is supposed to come between $D_c^{(l)}$ and $D_c^{(u)}$. Otherwise, the result is abnormal and is unacceptable for spatial analysis.

# 3 Empirical analysis

## 3.1 Study area, data, and method

A simple case is employed to make an empirical analysis to support the theoretical derivation and to demonstrate the application method. The study area covers the whole Chinese Mainland. The main geographical elements include 31 major cities in China in 2020 ($n$=31). The variables include six air pollution indicators of cities in China ($m$=6). Data came from *China Statistical Yearbook* in 2021. The statistical data quality may not be satisfactory, but for illustration and verification of a new method, the problem is not significant. Euclidean distance is utilized to define the generalized geographical space. Equations (13) and (15) are the key models for this empirical analysis. However, in empirical research, for simplicity, spatial correlation function can be equivalently replaced by spatial correlation number, which is as below:

$$N(r) = N^2 C(r) = \sum_{i=1}^{N}\sum_{j=1}^{N} H(r - d_{ij}), \quad (19)$$

where $N(r)$ refers to the correlation number based on the yardstick $r$. If $N(r)$ is treated as a cumulative correlation number function, then, $C(r)$ can be regarded as a cumulative correlation density function. Thus the generalized spatial correlation dimension can be estimated by the following power-law relation

$$N(r) = N_1 r^{D_c}, \quad (20)$$

in which $N_1$ denotes a proportionality coefficient. In theory, $N_1$ equals 1, but in practice, this parameter is not necessarily equal to 1. In practice, the $N_1$ value may deviate significantly by 1.

The preliminary processing of data can be completed through Microsoft Excel, and the calculation of correlation dimension can be realized by Matlab or other mathematical software. The calculation steps are as follows. First, standardize variables by $Z$-score. The so-called $Z$-score is a variable minus its average value and then divided by its standard deviation. Second, determine a generalized distance matrix by a proper formula. Euclidean distance can be calculated by equation (3), Minkowski distance can be calculated by equation (4), and Mahalanobis distance can be calculated by equation (5). Third, calculate spatial correlation numbers. The cumulative correlation are computed by using equation (19). Fourth, fit the fractal model to the dataset of correlation numbers. Equation (20) represents the fractal model for general correlation analysis. Fifth, examine and test the result of data fitting. If the relationships between the correlation number and correlation length follow a power law, the power exponent will give spatial correlation dimension, and we can make fractal dimension analysis. Otherwise, we should give up fractal method in this study, and return to the conventional mathematical methods. By the way, according to equation (19), it is easy to turn the correlation number function to correlation density function. If so, equation (19) should be replaced by equation (13). Accordingly, equation (20) should be replaced by equation (15). For a given yardstick length *r*, a correlation cumulative function changes to a correlation number $N(r)$.

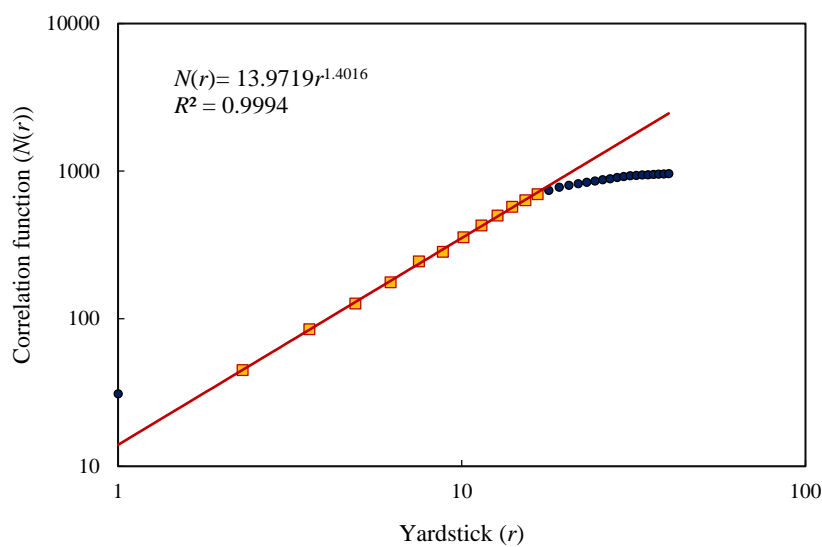

**Figure 2 Spatial correlation pattern of the major cities in Chinese Mainland based on longitude and latitude**

**Note:** All data points are represented by dots, and the scaling range is marked by squares. The squares may cover the dots. Fractal dimension is evaluated through scaling range. The equivalent relation between correlation function $C(r)$ and correlation number is $N(r)$ is $N(r) = NC(r)$. The same below.

In practical research work, spatial correlation is a local process rather than a global process. That is to say, there is a scaling range in a double logarithmic plot for fractal dimension estimation. If spatial scale is too small, there is no correlation object; in contrast, if spatial scale is too large, the correlation strength will be too weak to be taken into account. Many power laws can be treated as correlation functions. Therefore, power law usually break down when scale is too small or too large (Bak, 1996). The effective scale range of power law is called scaling range (Wang and Li, 1996). The so-called scaling range is the straight segment in a log-log plot. As a preparation, it is advisable to calculate the correlation dimension of urban distribution in the real geographical space of Chinese Mainland. Longitude and latitude are used to compute Euclidean distance, and correlation dimension can be easily estimated based on the distance matrix. The mathematical models for estimating correlation dimension of real space are exactly the same as equations (19) and (20). There is a clear scaling range in the log-log plot about the relationship between yardsticks and correlation function (Figure 2). The slope of the scaling straight line gives the estimated value of fractal dimension, $D$=1.4016. This result can provide reference for the subsequent correlation dimension analysis of generalized space.

**3.2 Calculation results based on original variables**

Five types of distances, including Chebychev distance, city block distance, Euclidean distance, Mahalanobis distance, and Minkowski distance, are employed to measure the correlation dimension in the generalized space of Chinese main cities based on urban pollution indicators. In order to lessen the impact of the problems of dimensions (quantitative units in variables), all variables are standardized using $Z$-score. First of all, let's examine the spatial correlation dimension based on Minkowski distance due to the generality of this distance. On log-log plot, the relationship between yardstick and correlation function shows a local straight linear trend. That is, there is scaling range in the scatterplot. The slopes of log-log straight line segments gives the estimated values of fractal dimension (Figure 3). Changing the parameter $p$ values yield different calculation results. The fractal dimension $D$ goes up slowly with the increase of parameter $p$. The process bears slight fluctuation. Gradually, the fractal dimension value converges to a constant. The final result is about $D$=1.79 (Table 1).

Table 1 Correlation dimension of the major cities in Chinese Mainland defined in generalized space based on four types of distances

| Type | Parameter | | Statistic | Scaling range | |
|---|---|---|---|---|---|
| | Coefficient $N_1$ | Dimension $D$ | Goodness of fit $R^2$ | Starting point | Point number |
| **City block distance** | 24.5791 | 1.5971 | 0.9971 | 2 | 12 |
| **Euclidean distance** | 71.6054 | 1.6625 | 0.9939 | 2 | 14 |

| | | | | | | |
|---|---|---|---|---|---|---|
| **Minkowski distance** | *p*=1 | 24.5791 | 1.5971 | 0.9971 | 2 | 12 |
| | *p*=2 | 71.6054 | 1.6625 | 0.9939 | 2 | 14 |
| | *p*=3 | 88.1923 | 1.7716 | 0.9961 | 3 | 15 |
| | *p*=4 | 99.9604 | 1.8104 | 0.9972 | 3 | 15 |
| | *p*=5 | 109.9392 | 1.7766 | 0.9940 | 3 | 18 |
| | *p*=6 | 112.4564 | 1.7893 | 0.9940 | 3 | 18 |
| | *p*=10 | 118.9693 | 1.7876 | 0.9941 | 3 | 18 |
| | *p*=100 | 122.9819 | 1.7873 | 0.9940 | 3 | 18 |
| **Chebychev distance** | | 122.9819 | 1.7873 | 0.9940 | 3 | 18 |

**Note:** "Starting point" implies the rank of the first data point falling into scaling range, and "point number" means the number of all data points falling into scaling range. The same below.

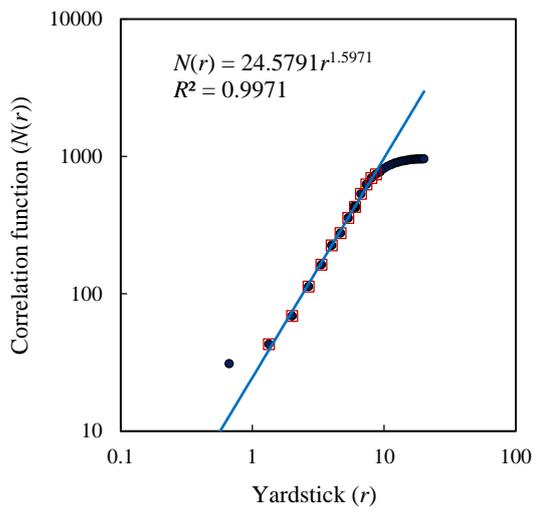

$N(r) = 24.5791 r^{1.5971}$
$R^2 = 0.9971$

a. Minkowski (1)

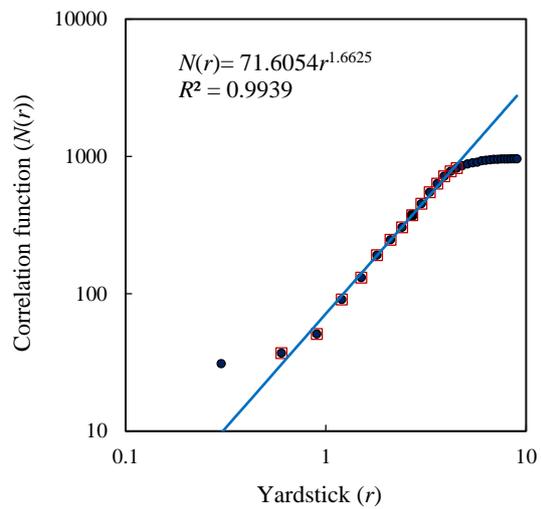

$N(r) = 71.6054 r^{1.6625}$
$R^2 = 0.9939$

b. Minkowski (2)

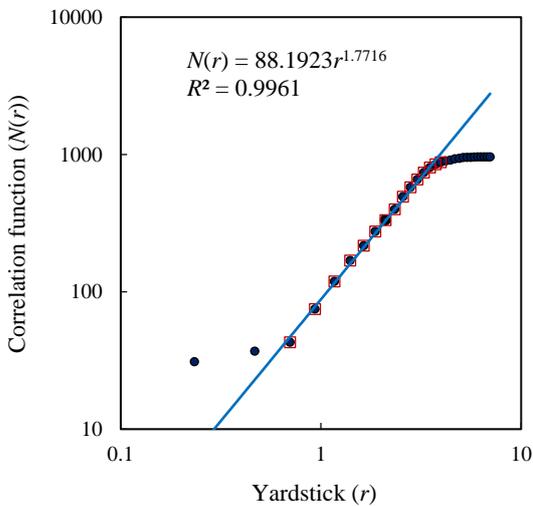

$N(r) = 88.1923 r^{1.7716}$
$R^2 = 0.9961$

c. Minkowski (3)

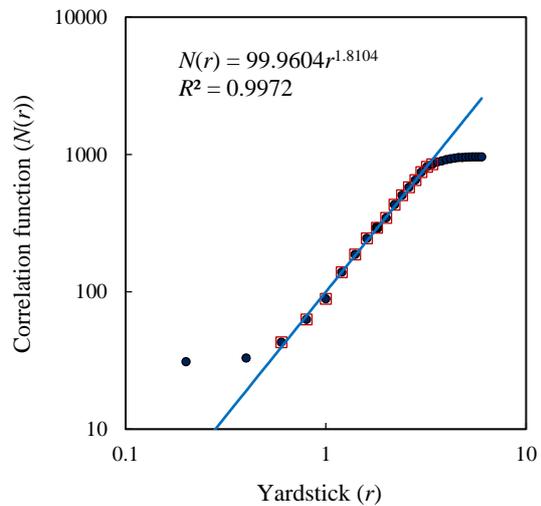

$N(r) = 99.9604 r^{1.8104}$
$R^2 = 0.9972$

d. Minkowski (4)

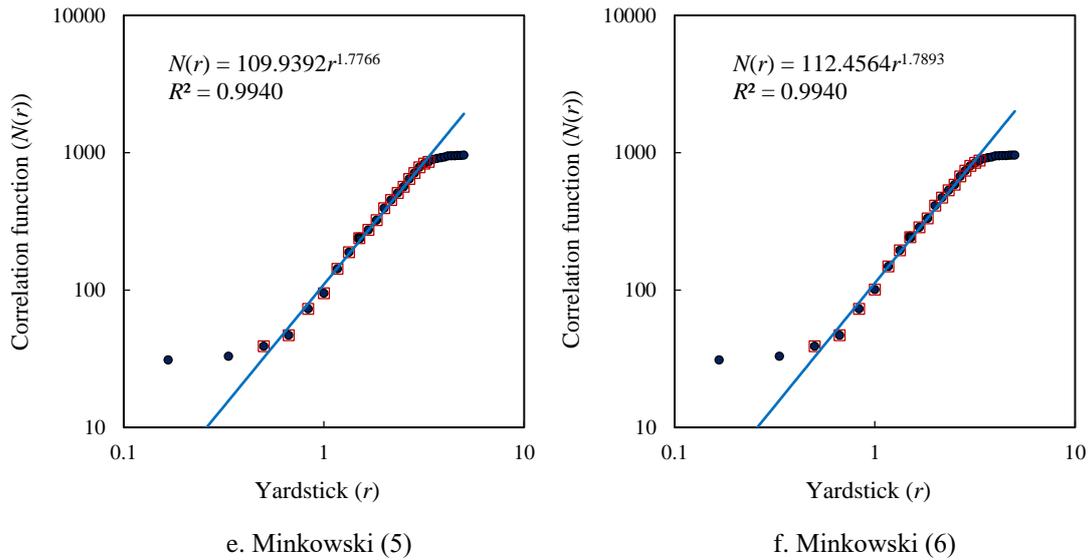

e. Minkowski (5)      f. Minkowski (6)

**Figure 3 The generalized spatial correlation patterns of the major cities in Chinese Mainland in based on Minkowski distance**

**Note:** The parameter value for Minkowski distance is $p$=1, 2, 3, 4, 5, 6, respectively. For $p$=1, the result is equivalent to city block distance; for $p$=2, the result is equivalent to Euclidean distance; for $p\to\infty$, the result is equivalent to Chebychev distance.

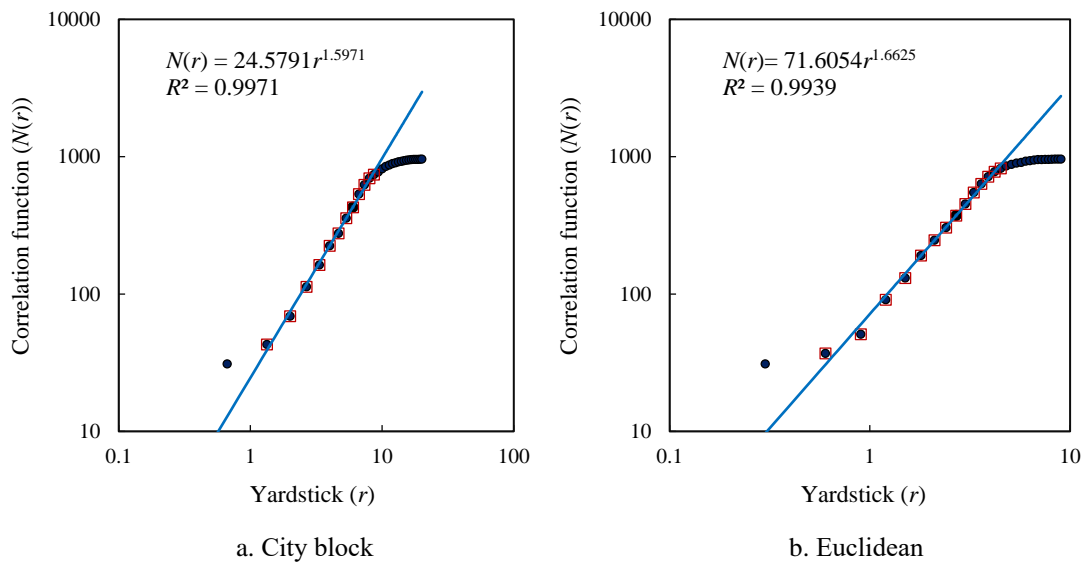

a. City block      b. Euclidean

**Figure 4 The generalized spatial correlation pattern based on city block and Euclidean distance**

As indicated above, Minkowski distance contains three special cases. If the parameter $p$=1, we have city block distance; If the $p$=2, we have Euclidean distance; If the $p\to\infty$, we have Chebychev distance. The fractal dimension value based on city block distance is about $D$=1.5971, which equals the fractal dimension value based on Minkowski distance when $p$=1; the fractal dimension value based on Euclidean distance is about $D$=1.6625, which equals the fractal dimension value based on Minkowski distance when $p$=2 (Figure 4). The fractal dimension based on Chebychev distance is around $D$=1.7873. Infinity is not a number, so a parameter with large value can be used to replace infinity. In fact, if $p$=100 as given, the fractal dimension value based on Minkowski distance is equal to the fractal dimension value based on Chebychev distance (Figure 5).

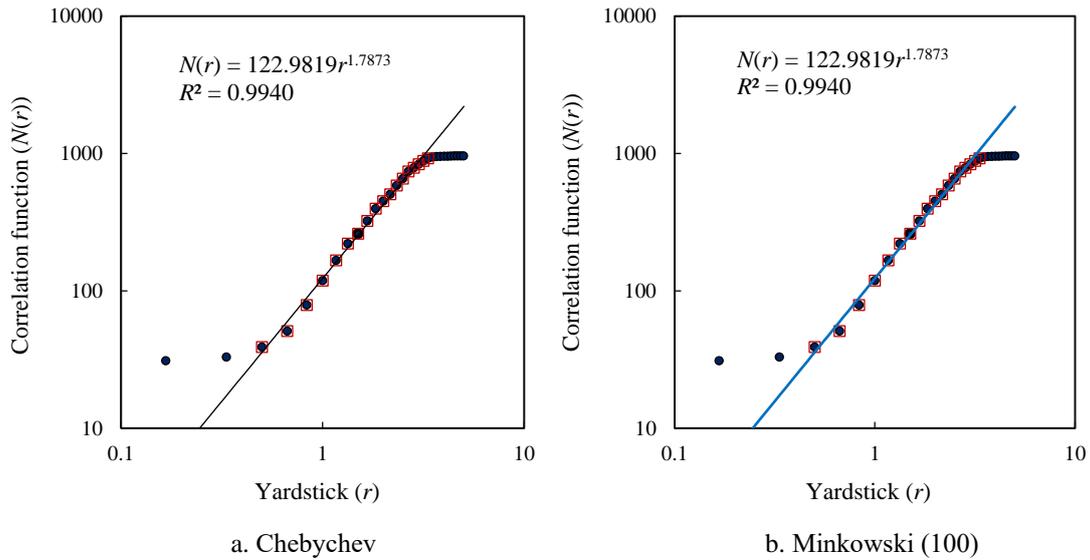

a. Chebychev     b. Minkowski (100)

**Figure 5 The generalized spatial correlation pattern of the major cities in Chinese Mainland based on Chebychev distance**

### 3.3 Calculation results based on factors

In quantitative research, there are two problems which significantly affect calculation results and analysis conclusions. One is inconsistency of dimensions (quantitative units), and the other is collinearity of variables. The consistency of dimensions and orthogonality of variables are preconditions of effective quantitative analysis. In the fractal dimension calculation in the previous subsection, variables have been standardized to reduce the impact of dimensions. However, the problem of collinearity between variables is not eliminated. One of the solutions to the variable collinearity problem is to use Mahalanobis distance. This is a type of generalized distance measure proposed by Mahalanobis (1936). It can be proved that Mahalanobis distance based on standardized variables is equal to the Euclidean distance base on standardized principal component scores, and standardized principal component scores can be understood as factor scores. Principal components represent new orthogonal variables transformed from standardized original variables. Therefore, there is no collinearity between any two principal components.

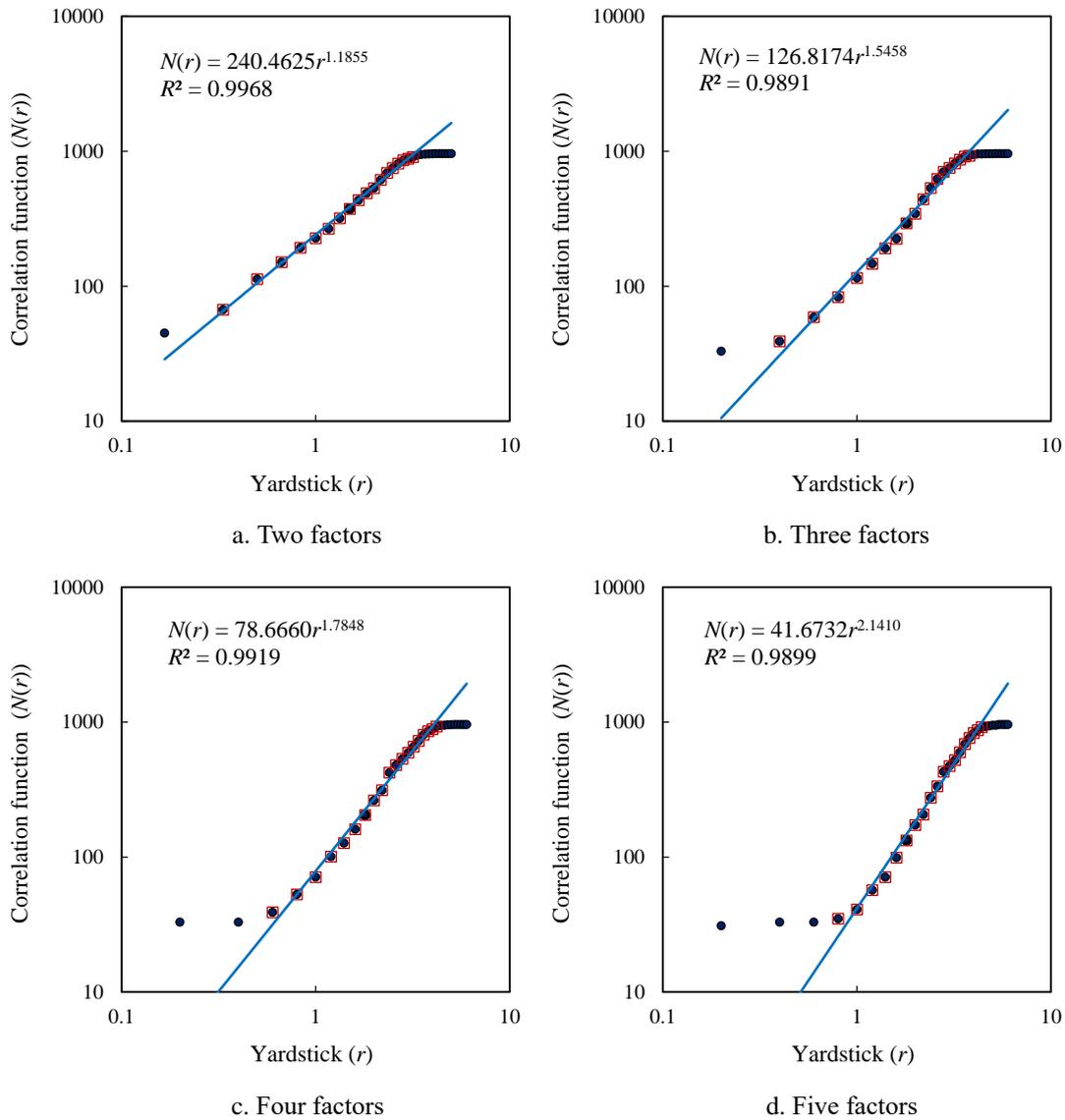

**Figure 6 The generalized spatial correlation patterns of the major cities in Chinese Mainland based on factor analysis**

**Note:** Euclidean distance matrix was obtained by two factors, three factors, four factors, and five factors, respectively. The correlation dimension was estimated by the Euclidean distance matrixes based on factor scores.

If the standardized variables is directly used to calculate the Mahalanobis distance, it is equivalent to using all factors to calculate the Euclidean distance. However, if all factors are taken into account, the unimportant factors will bring less useful information than the noise interference. Therefore, if necessary, we should discard the unimportant factors. Principal component analysis (PCA) was employed to extract factors. Each factor can be regarded as a standardized principal component. The importance of factors can be judged according to the variance contributions. The variance contributions of a factor load matrix equal the eigenvalues of standardized covariance matrix, and the eigenvalues equal the variances of principal components. We calculate Euclidean distances and generate distance matrixes by means of two factors, three factor, four factors, five factor, and six factors. Using each distance matrix, we can compute a correlation dimension. The results show that there are scaling ranges on the log-log plot of the relationships between yardsticks and correlation numbers. The slopes of the scaling range gives the estimated values of correlation dimension (Figure 6). The fractal dimension value increases with the number of factors (Table 2). If two factors are used, the fractal dimension value is close to 1; if more than 5 factors are used, the fractal dimension value exceeds 2. When six factors are used, the estimated value of fractal dimension is exactly the same as that of Mahalanobis distance (Figure 7). The results show that Mahalanobis distance is not suitable for correlation dimension analysis. The acceptable fractal dimension values come between 1 and 2 because that the Euclidean dimension of embedding space is $d_E=2$. With each factor added, the variance contribution rate will increase, but the added numerical value will become less and less significant. For this example, two factors can retain more than 80% of the information of the original variables, and three factors can retain about 90% of the information of the original variables (Table 3).

**Table 2 Correlation dimension of generalized space based on Euclidean distance of factors and Mahalanobis distance of original variables**

| Type | Factor number | Parameter | | Statistic | Scaling range | |
|---|---|---|---|---|---|---|
| | | Coefficient $N_1$ | Dimension $D$ | Goodness of fit $R^2$ | Starting point | Starting point |
| Euclidean distance based on factor | 2 | 240.4625 | 1.1855 | 0.9968 | 3 | 18 |
| | 3 | 126.8174 | 1.5458 | 0.9891 | 3 | 18 |
| | 4 | 78.6660 | 1.7848 | 0.9919 | 4 | 19 |
| | 5 | 41.6732 | 2.1410 | 0.9899 | 5 | 19 |
| | 6 | 25.8792 | 2.3582 | 0.9914 | 6 | 19 |
| Mahalanobis distance | 6 | 25.8792 | 2.3582 | 0.9914 | 6 | 19 |

**Note:** (1) The matrix of Euclidean distance of all factors is the same as the matrix of Mahalanobis distance of original variables. (2) The factor orthogonal rotation does not change the Euclidean distance matrix, thus does not change the fractal dimension values. (3) The fractal dimension value based on Mahalanobis distance is unreasonable.

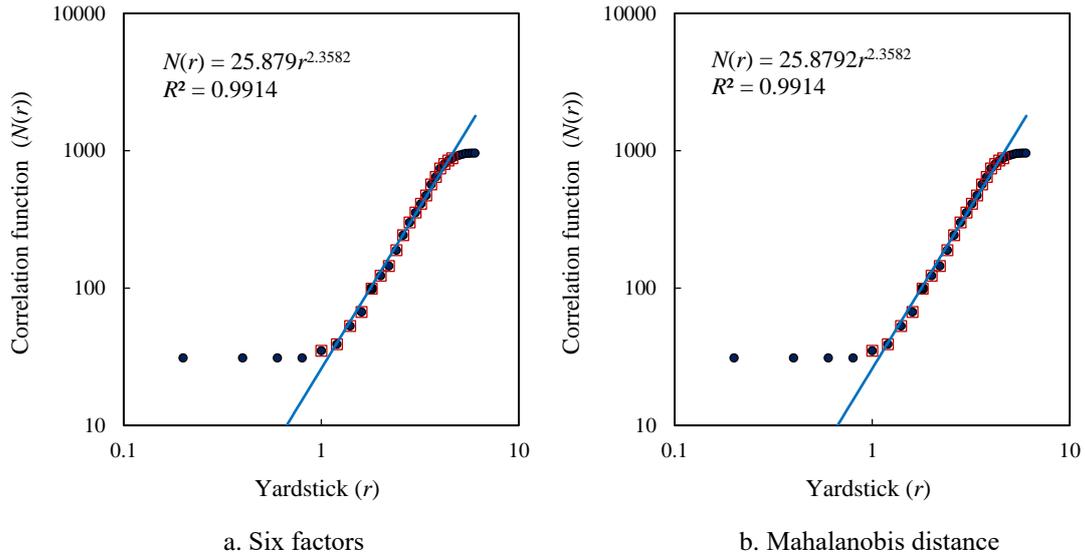

a. Six factors　　　　　　　　　　b. Mahalanobis distance

**Figure 7 The generalized spatial correlation patterns of the major cities in Chinese Mainland based on all factor analysis and Mahalanobis distance**

**Note:** If the number of factor *p* is equal to the number of original variables *m*, the Euclidean distances of factors equal Mahalanobis distances of original variables. Figure 6(a) is based on the Euclidean distance matrix of six factors, and Figure 6(a) is based on Mahalanobis distance matrix worked out from original variables.

### 3.4 Fractal features of generalized geo-space

Generalized geographical space is defined by mathematical distance or statistical distance rather than actual distance or traffic mileage. The distance matrix can be converted into a generalized map by multidimensional scaling analysis or similar methods (Haggett, 2001; Haggett *et al*, 1977). In the fractal research of geographical systems, distance matrix can be treated as one of starting points of fractal dimension estimation (Chen, 2008). Based on above empirical results, the main fractal properties can be outlined as follows. First, fractal structure appears within certain scaling range. In other words, there is scale-free range in a double logarithmic plot for the relationship between threshold distance (yardstick $r$) and correlation function (cumulative number $N(r)$ or density distribution $C(r)$). The slope of the straight line segment within the scaling range gives fractal dimension estimation values. If distance is too far, spatial correlation between a geographical element and other elements will be too weak so that the association can be ignored. On the other hand, if the distance is too close, no other geographical elements can be found, and the correlation between one element and itself will be extremely prominent. Consequently, there is a spatial range for fractal development, and the spatial range is often reflected by a scaling range. Second, fractal dimension values depend on distance definitions. This question is easy to understand. Based on different definitions of distance, the spatial pattern and structure reflected by distance matrix are different. Thus fractal dimension values are different. In practical studies, a comparable distance definition should be selected according to concrete research objective. Third, fractal dimension values depend on the number of factors. If multiple variables are used in a system analysis at the same time, the inter-variable multi-collinearity is inevitable. Collinearity leads to information affinity between variables, which leads to information redundancy. Multivariable information redundancy results in biased inferences or conclusions. One solution is to convert variables into factors, which can be obtained by principal component analysis. However, after variables are converted into factors, the redundant information between variables does not disappear, but is stored in the form of noise in the unimportant factors. The greater the variance of a principal component, the more useful information it carries about the original variable; on the contrary, the smaller the variance of a principal component, the less useful information it carries than useless noise. As indicated above, factors are standardized principal components. Mahalanobis distance is equivalent to the Euclidean distance based on all factors. In fractal dimension measurement, it is reasonable to use a few factors with large variance contributions instead of all factors.

**Table 3 Eigenvalues of covariance matrix and total variance explained**

| Factor | Initial eigenvalues | | | Extraction sums of squared loadings | | |
|---|---|---|---|---|---|---|
| | Total | % of Variance | Cumulative % | Total | % of Variance | Cumulative % |
| 1 | 3.9425 | 65.7077 | 65.7077 | 3.9425 | 65.7077 | 65.7077 |
| 2 | 1.1299 | 18.8320 | 84.5397 | 1.1299 | 18.8320 | 84.5397 |
| 3 | 0.3272 | 5.4534 | 89.9931 | 0.3272 | 5.4534 | 89.9931 |
| 4 | 0.3099 | 5.1647 | 95.1578 | 0.3099 | 5.1647 | 95.1578 |
| 5 | 0.2229 | 3.7148 | 98.8726 | 0.2229 | 3.7148 | 98.8726 |
| 6 | 0.0676 | 1.1274 | 100.0000 | 0.0676 | 1.1274 | 100.0000 |

As a case study, the ideas from fractals can be utilized to analyze the spatial properties and features of air pollution of 31 main cities in China. First, fractal structure suggests spatial association of air pollution of Chinese cities. The air pollution in different cities has the same cause or is related to each other. Otherwise, it is impossible to form fractal structure in the generalized geographical space. Second, fractal dimension values suggest that spatial correlation of urban pollution is strong. As indicated above, correlation dimension is defined by Renyi entropy and correlation function. Based on different definition of distance, valid correlation dimension values in the generalized space vary from 1.55 to 1.79. However, the correlation dimension of urban systems in the real space is about 1.4 (Figure 2). This means that the correlation of air pollution is much stronger than that of cities in the real world through traffic and communication network. Third, scaling range suggests that spatial correlation of urban air pollution are significant only within certain distance. The distance in the generalized space is related to the distance in the real space. A distance matrix can be converted into a map by multidimensional scaling analysis. By means of scaling range, we can judge the correlation length of urban air pollution.

## 4 Discussion

The basic metric of geo-spatial analysis is distance, including real distance, mathematical distance, and statistical distance. All these metrics can be reflected by Euclidean distance, that is, linear distance. For real geographical space, we can describe it using linear distance or traffic mileage between geographical elements. In fractal studies, the linear distance is termed crow distance, and the traffic mileage can be measured by road distance and railway distance and termed cow distance (Kaye, 1989). It is easy to calculate the fractal dimension of the real geographical space by varied methods (Batty and Longley, 1994; Frankhauser, 1998b; Takayasu, 1990). For generalized geographical space, we can only characterize it by means of mathematical distance or statistical distance. These distances can be measured by a number of observed variables of a set of geographical elements ($m$ variables, $n$ elements, $m \ll n$). Using the geographical distance, we can define a spatial correlation function. Fitting the generalized spatial correlation function ($C(r)$ or $N(r)$) to observational data yields certain statistic relationships. If the statistical relationships between yardsticks $r$ and correlation function $C(r)$ or $N(r)$ do not follow a power law, it indicates that the generalized geographical space bear characteristic scale and can be analyzed by conventional mathematical methods. In this case, no fractal property can be found. On the contrary, if the statistical relationships follow a power law, it suggests fractal structure in the generalized geographical space. Then, we can employ the ideas from fractals to make spatial analysis.

Correlation dimension is one of parameters for describing multifractal systems. In the multifractal spectrums of generalized correlation dimensions, there are three basic parameters, that is, capacity dimension, information dimension, and correlation dimension (Grassberger, 1983). As far as fractal urban systems are concerned, the meanings of the three important multifractal parameters are as follows (Table 4). It is easy to measure and compute capacity dimension and information dimension of real geographical space. How can we calculate the capacity dimension or information dimension for generalized geographical space? A finding is that hierarchical cluster analysis can employed to estimate capacity dimension of abstract geographical space based on multivariable. For the case discussed in Section 3, the hierarchical clusters analysis can be made by means of standardized Euclidean distance and furthest neighbor method to estimate fractal dimension value. The capacity dimension of the 31 major cities in Chinese Mainland defined in generalized space based on six variables is about $D$=1.5855 (Figure 8). Further, a similar method may be developed to calculate the information dimension of cities in the generalized geographical space.

**Table 4 The spatial meanings of three basic multifractal parameters for complex urban systems**

| Fractal parameter | Symbol | Variable | Meaning |
|---|---|---|---|
| **Capacity dimension** | $D_0$ | Categorical variable (Boolean numbers) | Is there a city in a place? Yes, it is 1. No, it is 0. |
| **Information dimension** | $D_1$ | Metric variable (probability) | If there are cities in a place, what is the proportion of the number of cities in the whole region? |
| **Correlation dimension** | $D_2$ | Categorical or metric variable | If there is a city in one place, can we find other cities within a given distance? |

**Note**: Categorical variable is also termed dummy variable or indicator variable or nominal variable.

The novelty of this work rest with the mathematical derivation and empirical analysis of correlation dimension of generalized geographical space. New models and analytical framework for fractal systems based on multivariate random variables were developed. Traditional clustering analysis is tentatively used to estimate the fractal dimension of abstract space, and this may inspire new directions for the joint application of scaling analysis and traditional mathematical methods. In previous studies, spatial correlation modeling was researched for real geographical space and hierarchical order space (Chen, 2008; Chen, 2013; Chen, 2014; Chen and Jiang, 2010). A phase space can be reconstructed by means of nonlinear time series analysis based on Takens' theorem (Kantz and Schreiber, 2004; Takens, 1981). The method of phase space reconstruction can be used to compute correlation dimension in time series of geographical evolution. In this paper, the correlation dimension analysis is generalized to more abstract geographical space. Thus, geographical space can be divided into four categories (Table 5). If an appropriate distance is defined for generalized geographical space and a distance matrix is gained, we can compute correlation dimension and make spatial correlation analysis. In this way, the application sphere of fractal geometry in geographical analysis will be greatly expanded. The shortcomings of this study lies in two aspects. First, the visualization problem of generalized geospatial has not been solved yet. Generalized geographical space is abstract, and thus is not easy for readers to understand intuitively. Only by visualizing generalized space can this analysis method be better applied to practical problems. Second, the empirical analysis is not in-depth. This paper is devoted to developing geographical fractal theory and analytical method rather than positive analysis. The case study is used to support the theoretical derivation results and demonstrate the calculation method of generalized correlation dimension. Pure and in-depth case analysis based on generalized correlation dimension will be carried out for complex geographical systems in the future.

Table 5 Geospatial classification based on spatial correlation dimension

| Space type | Data | Method | Fractal dimension |
| --- | --- | --- | --- |
| **Real geo-space** | Maps or remote sensing images | Spatial measurement | Spatial correlation dimension |
| **Generalized geo-space** | Time series | Reconstructing phase space | Temporal correlation dimension |
| | Cross-sectional series | Reconstructing hierarchical space | Hierarchical correlation dimension |
| | Mathematical or statistical distance | Constructing metric space based on multi-variables | Generalized spatial correlation dimension |

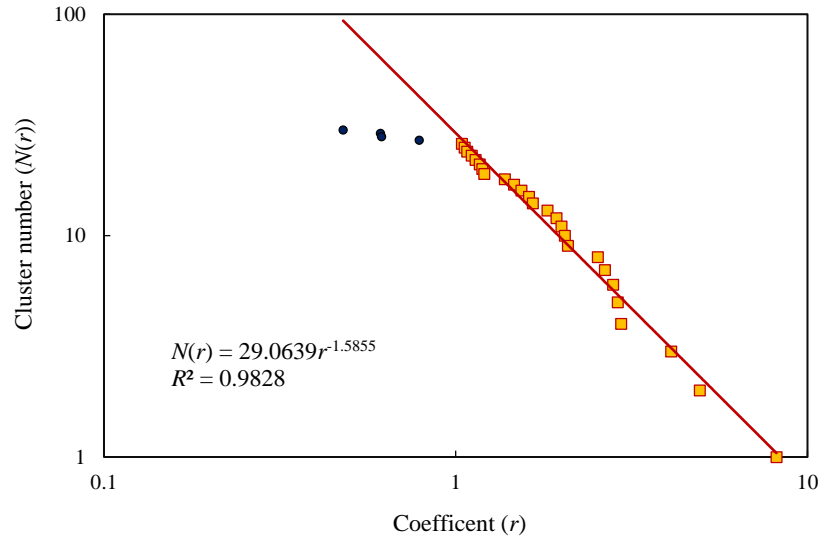

**Figure 8 Hierarchical cluster analysis is used to estimate the capacity dimension of the major cities in Chinese Mainland defined in generalized space**

**Note:** (1) The cluster analysis is based on standardized Euclidean distance and furthest neighbor. (2) All data points are represented by dots, and the scaling range is marked by squares.

# 5 Conclusions

Geographical space falls into two categories: real geographical space and generalized geographical space. The latter can be defined by mathematical or statistical distance. Generalized geographical space can be divided into two types, one is that with characteristic scale and the other is that without characteristic scale. If a generalized geographical space bears characteristic scale, it can be modeled and analyzed by conventional mathematical methods. If a generalized geographical space bears not characteristic scale, it can be researched by using fractal geometry. The main conclusions can be reached as follows. *First, spatial correlation dimension analysis can be applied to scale-free generalized geographical space defined by a set of random variables*. As long as an appropriate distance definition is found, a generalized geographical space can be constructed and the spatial correlation numbers can be calculated. If the relationship between spatial correlation function and threshold distance obeys a power law, it can be considered that it has no characteristic length, then the generalized correlation dimension analysis can be carried out for obtaining useful spatial information. If the correlation function does not obey the power law, it shows that it has a characteristic length, so the fractal method is invalid and can only be modeled by conventional mathematical methods. *Second, spatial correlation function has a scaling range, in which the correlation dimension can be effectively estimated*. Spatial association is not infinite. If the scale is too small, that is, the distance is too close, there may be no associated objects. On the contrary, if the scale is too large, that is, the distance is too far, the correlation may be too weak to be considered. Therefore, only in a certain scale range, the correlation function shows a power-law relationship. Correspondingly, in a log-log plot, only within the scaling range can a clear straight line segment be formed. The slope of the straight line segment indicates the generalized correlation dimension. *Third, spatial correlation dimension values depend on definitions of mathematical or statistical distance*. The distance matrixes based on different definitions give different spatial patterns of geographical systems. So the correlation dimension values based on different distance definitions are different to each other. It is necessary to determine a proper and comparable distance definition according to the properties of geographical systems and research objectives. In particular, the dimensional homogeneity and orthogonality of variables are two prerequisites of effective distance calculation. Therefore, selecting appropriate variables and unifying dimensions (quantitative units) are necessary conditions for generalized correlation dimension analysis.

**Acknowledgement**

This research was sponsored by the National Natural Science Foundation of China (Grant No. 42171192). The support is gratefully acknowledged.# References